\documentclass[conference]{IEEEtran}

\AtBeginDocument{%
  \providecommand\BibTeX{{%
    \normalfont B\kern-0.5em{\scshape i\kern-0.25em b}\kern-0.8em\TeX}}}

\usepackage{multirow}
\usepackage{booktabs}
\usepackage{color, colortbl}
\usepackage{placeins}
\usepackage{graphicx}
\usepackage{cite}
\usepackage{hyperref}
\usepackage{balance}

\begin{document}

\title{Exploring Privacy Implications\\in OAuth Deployments}

\author{\IEEEauthorblockN{Srivathsan G. Morkonda, Paul C. van Oorschot and Sonia Chiasson}
\IEEEauthorblockA{\textit{School of Computer Science, Carleton University} \\
\textit{Ottawa, ON, Canada} \\
\textit{Email: srivathsan.morkonda@carleton.ca, paulv@scs.carleton.ca, chiasson@scs.carleton.ca}}}

\maketitle

\begin{abstract}
Single sign-on authentication systems such as OAuth 2.0 are widely used in web services. They allow users to use accounts registered with major identity providers such as Google and Facebook to login on multiple services (relying parties). These services can both identify users and access a subset of the user's data stored with the provider. We empirically investigate the end-user privacy implications of OAuth 2.0 implementations in relying parties most visited around the world. We collect data on the use of OAuth-based logins in the Alexa Top 500 sites per country for five countries. We categorize user data made available by four identity providers (Google, Facebook, Apple and LinkedIn) and evaluate popular services accessing user data from the SSO platforms of these providers. Many services allow users to choose from multiple login options (with different identity providers). Our results reveal that services request different categories and amounts of personal data from different providers, with at least one choice undeniably more privacy-intrusive. These privacy choices (and their privacy implications) are highly invisible to users. Based on our analysis, we also identify areas which could improve user privacy and help users make informed decisions.
\end{abstract}

\section{Introduction}
An increasing number of web applications encourage users to log in to their services in exchange for a personalised experience. However, this comes with a usability problem for users having to administer a growing list of account credentials, e.g., to choose unique and strong passwords for accounts on each service. Managing large sets of credentials is a difficult task for users and can result in insecure practises such as reusing passwords and choosing weak passwords~\cite{stobert2014password}. Many web authentication schemes have been proposed to improve usability and convenience. Federated single sign-on (SSO) schemes involve a trust relationship between an \textit{identity provider} (IdP) and one or more other-party services (\textit{relying parties} or RPs) that allow users to identify themselves on the service using login credentials registered with the IdP. The OAuth 2.0 protocol is an example of a federated SSO scheme used to establish trust in identity-related interactions between an IdP and an RP.

Instead of requiring users to create new credentials, RPs can use the OAuth 2.0 framework~\cite{oauth2rfc} to outsource user identification to an IdP with whom users are likely to have an existing account. As part of this transaction, the RP can also request access to additional user personal data stored with the IdP. Major identity providers (e.g., Facebook, Google, Microsoft) expose web APIs to grant RPs controllable access to protected user data stored on their platforms. With the user's permission, an IdP allows the RP access to one or more user-data attributes, which in some cases includes sensitive user data such as emails, contact information and documents in personal cloud storage---raising privacy concerns. Such user data allows the RP to extend functionality and personalise web content to the user. It also reduces implementation costs for the RP since they are outsourcing login-related tasks, including key management and credential verification, to the identity provider. 

\begin{figure*}[tb]
    \centerline{\includegraphics[width=.8\textwidth]{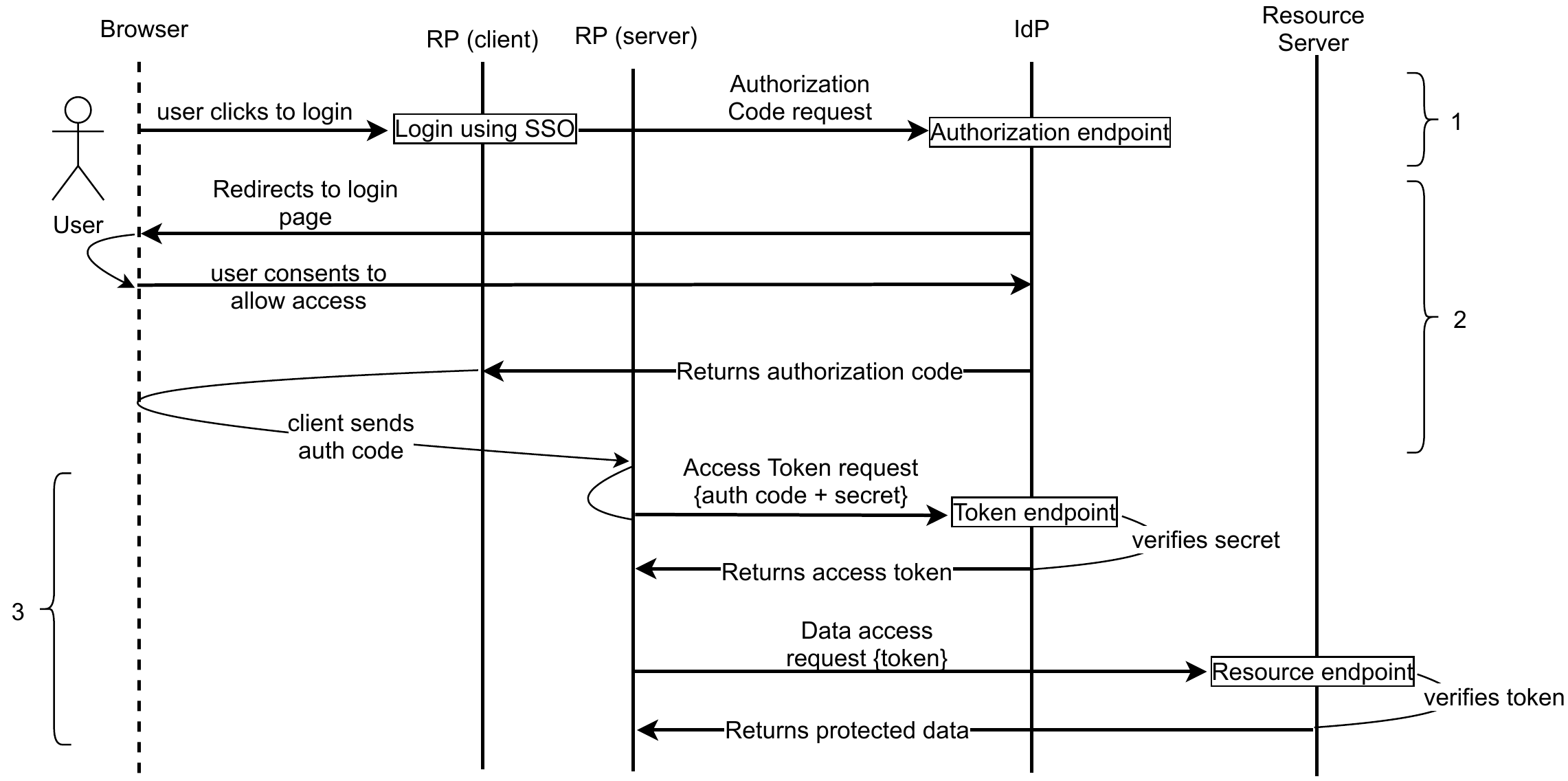}}
    \caption{Procedure for OAuth 2.0 Authorization Code flow (based on~\cite{oauth2rfc}). Diagram for the Implicit flow is in Appendix~\ref{app.implicit}.}
    \label{figAuthCodeFlow}
\end{figure*}

Prior work (e.g., ~\cite{drakonakis2020cookie}~\cite{ghasemisharif2018single}) involving OAuth-based SSO gives focus to security issues including leaks of user data from relying parties to other third-party actors due to implementation flaws. In contrast, we focus on privacy consequences for users of OAuth-based SSO that arise when RPs gain access to user data stored on IdP sites. In this study, we evaluate privacy practises in popular services using OAuth for SSO. Our contributions include:
\begin{itemize}
    \item \textit{OAuthScope}, a tool to extract OAuth protocol request parameters from sites supporting major SSO providers (Google, Facebook, Apple and LinkedIn).
    \item An empirical study of OAuth-based logins in the Alexa Top 500 sites for five countries. The study reveals considerable variation in how relying parties implement the different SSO options, in the data made available by identity providers, and in apparent trends across countries.
    \item An explication of how user choices, typically made without full information, can result in release of considerably different amounts of user data and have privacy implications.
\end{itemize}

\noindent
Based on our analysis, we further discuss possible modifications to OAuth which could lead to improved end-user privacy.

Note: while some enterprise SSO providers (e.g., Microsoft) are popular among users, these are primarily used in closed systems using enterprise accounts. Since we target websites involving personal user accounts, we do not include these providers in our study.

The next section provides background on OAuth 2.0 framework. Section~\ref{sec.oauthscope} introduces the OAuthScope tool. Section~\ref{sec.empirical.study} presents the empirical study. Section~\ref{sec.api.analysys} provides our classification of user data available in four identity providers. Results of the empirical study are reported in  Section~\ref{sec.results}. Section~\ref{sec.privacy.implications} presents our analysis of privacy implications to SSO users. In Section~\ref{sec.recommendations}, we propose stakeholder-based changes to improve user privacy. Related work is described in Section~\ref{sec.related.work} and concluding remarks in Section~\ref{sec.concluding.remarks}.

\section{OAuth 2.0 Framework (Background)}
\label{sec.framework}
OAuth 2.0~\cite{oauth2rfc} is a web resource authorization protocol popular in client-server deployments worldwide for granting applications access to protected resources without sharing the user's credentials. For example, a website prompts a user to log in and optionally allow access to specified user data by relying on their Google account rather than creating new credentials on that website. The website does not gain access to the user's Google credentials, but instead gains access to a subset of the user's data and trusts Google's verification of the user's credentials. The user is able to use the same Google account to log in on other websites that support Google as a SSO option. User data is referred as \textit{protected resources} and the user is called as the \textit{resource owner} (RO) in OAuth 2.0 specification.

The OAuth 2.0 protocol is composed of several ``grant types" that define how credentials are granted to RPs. The procedure for obtaining access to protected resources is defined by \textit{OAuth flows} and each flow is designed to serve different use cases (and provide different levels of security). Since our study involves OAuth 2.0 use in web applications, we describe the steps involved in two flows commonly used in these applications.

\subsection{Authorization Code Flow (Server-side flow)}
In order for an RP to use OAuth 2.0, it must first register itself with an IdP to obtain a \texttt{client\_id} (used for identification of the RP in requests). Additionally, in the case of confidential clients (RPs with the ability to securely store secrets), an associated \texttt{client\_secret} is issued by the IdP. This allows an IdP to authenticate requests from an RP. We provide an overview of the three primary steps (labelled in Fig.~\ref{figAuthCodeFlow}) involved in the \textit{authorization code flow}~\cite{oauth2rfc}.
\begin{enumerate}
\item \textbf{RP request to IdP}: The resource owner triggers the flow by clicking a login element, sending a request constructed by the RP to the IdP's \textit{authorization endpoint}. The RP programs the request to add several parameters (as query components in the request URI) specifying the access request. To indicate this flow type, the RP must set the value of \texttt{response\_type} to \texttt{code}. The request also includes the parameters \texttt{scope} and \texttt{redirect\_uri}. The \texttt{scope} parameter lists one or more protected resources (e.g., name, email) the RP is requesting to access. The allowed values for this parameter are specified by individual IdPs based on the resources made available by the IdP. The \texttt{redirect\_uri} is the endpoint the RP is requesting the IdP to redirect the RO, at the end of the flow. In order to redirect to the intended party, the OAuth 2.0 specification~\cite{oauth2rfc} requires the IdP to check if the URI specified in the request matches with the value registered by the RP.

\item \textbf{IdP request to RO}: After receiving the RP's request, the IdP redirects the RO to a login page on its domain to login with the IdP account. If the RO is already logged in, the IdP displays a prompt to confirm resource access to the RP. In this process, the IdP provides the RO with a list of user-data attributes the RP is requesting to access and optionally provides the ability to deny RP the access to one or all of the requested attributes. If the RO approves the request, the IdP issues an authorization \texttt{code} and redirects the RO back to the RP at the verified redirect endpoint. The \texttt{code} is included as a query component to the redirection URI. The standard does not specify what should happen if the RO denies access, leaving it to the IdP's discretion. 

\item \textbf{Token exchange}: From its server-side application (which is considered more secure than a channel from the RO's browser), the RP uses the \texttt{client\_secret} and the authorization \texttt{code}  to exchange for an \textit{access token}. After verifying the \texttt{code} and RP's \texttt{secret}, the IdP issues an access token allowing the RP to access RO's resources within the IdP. Access tokens can vary in format (chosen by each IdP), but the purpose is to represent authorization information (e.g., allowed scope values, token expiry date). A commonly used self-contained format is the JSON Web Token (JWT)~\cite{jwtrfc} that protects the integrity of information within the token.
\end{enumerate}

\subsection{Implicit Flow (Client-side flow)}
\label{sec.framework.implicit}
The \textit{implicit} grant type is simpler than the authorization code flow in that it allows the RP to directly obtain the access token without exchanging an authorization code. It is useful for browser-based (JavaScript) apps that lack the ability to securely store secrets. The RP initiates the implicit flow by sending an access request to the IdP with the parameter \texttt{response\_type=token}. The IdP prompts the user with a login screen and, if access is granted, an access token is issued and included in the URI fragment of the redirect endpoint. The redirect URI is verified by the IdP to ensure it matches the value registered by the RP. Since the access token is returned in the redirection URI, it is included in the user's browsing history and, therefore, the security of the access token relies on the security of the user's system. A malicious application with access to the user's browser could misuse the access token. Additionally, third-party scripts running within the RP site will be able to access the token. 

The OAuth 2.0 implicit flow was originally designed when browsers restricted apps to make requests only to its own domain~\cite{oktaImplicitDead}. This browser restriction prevented browser-based apps from using the authorization code grant since it requires sending a HTTP POST request to the IdP's authorization endpoint, which in many apps (not owned by the IdP) is different from the RP. The implicit flow in OAuth 2.0 offered a workaround that avoids using the POST request and includes the access token directly in the redirection URI. Though implicit flow provides the needed functionality, this is not recommended from a security perspective due to the risks associated with storing credentials in URIs. Modern browsers now support Cross-Origin Resource Sharing (CORS) that enables a website to request resources from other permitted origins, removing the need for this workaround. The challenge still remains for browser-based applications to securely store secrets, and the recommended flow for such situations is the authorization code flow with Proof Key for Code Exchange (PCKE)~\cite{oauth2pkce}.

\subsection{Authentication (OpenID Connect)}
\label{sec.authentication}
OAuth 2.0 can be adapted to allow an IdP to authenticate users to the RP. The OpenID Connect 1.0~\cite{openIdConnect1Spec} (OIDC) specification is designed for authentication and is built upon the OAuth 2.0 protocol. OIDC introduces a special value (\texttt{openid}) to the \texttt{scope} parameter for specifying intent to authenticate. The OIDC specification also extends OAuth 2.0's \texttt{response\_type} parameter to define additional flows. If this parameter includes the value \texttt{id\_token}, the OpenID Provider (OP) will issue an \textit{ID token} to the RP after a successful authentication, encoded as a JWT~\cite{jwtrfc} and containing key-value pairs (\textit{Claims}) about the user's identity. \textit{Claims} are digitally signed using JSON Web Signature (JWS). Since many RPs use both OIDC and OAuth 2.0, we refer to \textit{OpenID Providers} as IdPs in our study. OIDC allows any combination of \texttt{id\_token}, OAuth 2.0 values \texttt{code} and \texttt{token} for the \texttt{response\_type} parameter~\cite{openIdConnect1Spec}. Each combination refers to a \textit{hybrid flow} that defines the values (access token, authorization code and ID token) included in IdP response and the endpoint (authorization and token) issuing the values.

Although the OIDC standard is built on top of OAuth 2.0 specification, some identity providers instead use custom modifications of OAuth 2.0 to provide authentication capabilities. In Section~\ref{sec.api.analysys}, we briefly describe these modifications as part of our analysis of the four identity providers.

\subsection{Refresh Tokens}
\label{sec.framework.refreshTokens}
An OAuth 2.0 access token is issued for a limited lifetime and once it expires, the user is required to approve access again in order for a new access token to be issued to the RP. A common access token type is ``bearer" token that allows use by any party in possession of the token. An unauthorized party in possession of such tokens can use it to access protected resources. While long-lived access tokens improve user experience by reducing the frequency of required logins, the longer life increases the risk of token leaks. The recommended approach is to combine the use of a access token with a \textit{refresh token}, a string issued by the IdP that allows the RP to extend existing access without involving the user to approve access again. The RP can extend its access by exchanging the refresh token for a new access token with a scope (identical or lesser than the previously issued access token) determined by the IdP. During this exchange, the IdP validates the refresh token before issuing a renewed access token and optionally, a new refresh token. Due to the sensitivity of refresh tokens, the specification restricts its use to only server-side flows such as the authorization code flow~\cite{oauth2rfc}.

\section{The OAuthScope tool}
\label{sec.oauthscope}
We designed and built OAuthScope, a Java web tool that scans and extracts OAuth 2.0 protocol-related parameters from authorization requests made by relying parties to identity providers. We provide a list of URLs as input, and OAuthScope visits each URL in a headless browser setup from a local server. It scans each site to locate OAuth-based SSO requests to any of the four providers in our study (i.e., Google, Facebook, Apple, and LinkedIn). For each SSO option available on the RP, OAuthScope simulates a mouse-click to trigger an OAuth 2.0 flow, which initiates an authorization request by the RP to the IdP's authorization endpoint. OAuthScope captures the parameters included in the request, including the flow type and \texttt{scope} parameter that specifies the protected resources for which the application intends to obtain access. OAuthScope uses \textit{Selenium WebDriver}~\cite{seleniumWebDriver}, an open-source browser automation framework built primarily for automated in-browser testing of web applications. OAuthScope uses the \textit{WebDriver} framework for identifying HTML elements on an RP site and simulating web page navigation to extract OAuth 2.0 protocol data from supported IdP sites. We summarize the primary steps performed by OAuthScope as follows:

\begin{enumerate}
    \item \textbf{Identify login element}: After loading a website's landing page, OAuthScope searches for potential login elements based on a set of predefined match criteria and simulates a user click if an element is found. Most RPs display login forms in either a new page or a new iframe within the landing page. We consider both cases and switch contexts as necessary. Our tool captures a screenshot of the login page, for use in manual verification of correctness.
    \item \textbf{Identify SSO elements}: On the new page (or iframe), it performs a search for HTML elements that potentially lead to a login page of one of the four IdPs in our study. OAuthScope identifies SSO elements based on element texts and HTML tag values commonly found in RPs pointing to IdPs. After obtaining potential SSO login elements, our tool simulates clicks on each element and checks if the resulting page's URL matches with a list of predefined endpoints for each IdP in our study. We built this list to contain OAuth 2.0 endpoints published on each IdP's developer documentation pages.
    \item \textbf{Extract OAuth 2.0 parameters}: The OAuth 2.0 framework specifies that protocol parameters should be added as query components of the request URI (an example is listed in Appendix~\ref{app.authorization}) initiated by an RP. OAuthScope extracts these parameters encoded in the link as key-value pairs. Finally, the parameters and login screenshots are persisted to a database for further analysis.
\end{enumerate}

We note that the current version of OAuthScope does not automatically ensure identification of all SSO elements in all websites. A manual inspection of the web pages is still necessary to ensure completeness of the collected data. To facilitate this process, our tool captures screenshots of the login pages and of the landing page for any site where our tool did not find a login element, and stores the screenshots to a database. We use these images to identify any SSO options missed by the automated scan and manually feed the SSO login links to OAuthScope for extraction of the protocol parameters. Since our goal in this study is to evaluate privacy implications of top sites using OAuth-based SSO, we did not focus our efforts on building a tool that can fully automate the data extraction process. However, we did iteratively improve our match criteria based on strings and tags found in RP sites during our data collection process. This gradually reduced the need for manual involvement in data collection. OAuthScope includes a web front-end (included in Appendix~\ref{app.screenshot}) for displaying and analysing the collected dataset.

\section{Empirical Study: Overview}
\label{sec.empirical.study}
We conducted a study of SSO systems in popular websites that implement the OAuth 2.0 framework for identifying users and accessing user data stored with different identity providers. In an initial exploration, we manually scanned the Alexa Top 500 sites~\cite{alexaTopSites} in a first target country and found that three major identity providers (Google, Facebook and Apple) are predominantly supported as SSO options. In addition to these providers which primarily contain personal data of users, we also included LinkedIn as an IdP given its popularity as a platform for sharing professional data. 

For each of these four IdPs, we collected OAuth 2.0 protocol-related data from the Alexa Top 500 sites in five countries: Australia, Canada, Germany, India and the United States. We describe our data collection procedure below.

\subsection{Research Questions}
The goal of this study is to understand the privacy implications for users opting to use OAuth-based SSO to log in to the top websites. To achieve this, we pursue the following research questions:
\begin{itemize}
    \item [RQ1:] What categories of user data do relying parties request from SSO providers? (Sec.~\ref{sec.results.dataRequestsAcrossCountries},~\ref{sec.results.dataRequestsAcrossProviders})
    \item [RQ2:] How prevalent is each SSO scheme in popular relying parties across the five countries? (Sec.~\ref{sec.results.flows})
    \item [RQ3:] If a relying party supports multiple SSO options, how do they differ in terms of requested user-data attributes? (Sec.~\ref{sec.results.dataRequestsAcrossProviders})
\end{itemize}

\subsection{Data Collection}
To collect accurate representation of websites as served to local users in each country, we use a VPN service to connect to a server in the country when scanning and collecting data from sites. For each country included in our study, we first manually visit each of the Alexa Top 500 sites in the country and identify websites that support at least one of our four chosen IdPs. We then run each filtered site through OAuthScope to extract the OAuth 2.0 parameters for each of these IdPs supported by the website. As a cross-check, we noted the IdPs supported by each website during our initial manual scan and verified that OAuthScope collected all available data from each website. For any omissions, we manually obtained the IdP links and passed it to OAuthScope for extraction of the protocol parameters. In total, this provided us with a dataset consisting of details from 815 RPs  (Australia: 174; Canada: 159; Germany: 126; India: 172; US: 184). Our dataset consists of all RPs from each country that use at least one of the four IdPs; if a site appeared in the top 500 list of more than one country, it is included each time so that we could make direct comparisons across countries. 

\begin{table*}[]
\centering
\footnotesize
\caption{Comparison of data attributes in provider APIs. *Data not requested (but available) by any website in our dataset. Default fields are italicized. For further info. explaining the entries, see Google~\cite{googleOauth2Scopes}, Facebook~\cite{facebookOauthPermissions}, Apple~\cite{appleSignInDocumentation} and LinkedIn~\cite{signInWithLinkedIn}.}
\begin{tabular}{lllll}
\hline
\multicolumn{1}{l}{\textbf{Data category}} &
  \multicolumn{1}{l}{\textbf{Google}} &
  \multicolumn{1}{l}{\textbf{Facebook}} &
  \multicolumn{1}{l}{\textbf{Apple}} &
  \multicolumn{1}{l}{\textbf{LinkedIn}} \\ \hline
\textbf{Basic} &
  \begin{tabular}[c]{@{}l@{}}email\\ \textit{profile}\\ openid\end{tabular} &
  \begin{tabular}[c]{@{}l@{}}email\\ \textit{public\_profile}\end{tabular} &
  \begin{tabular}[c]{@{}l@{}}email\\ name\end{tabular} &
  \begin{tabular}[c]{@{}l@{}}r\_emailaddress\\ name\\ profilePicture\\ headline\end{tabular} \\ \hline
\textbf{Identity} &
  \begin{tabular}[c]{@{}l@{}}user.birthday.read\\ user.addresses.read*\\ user.gender.read*\\ user.phonenumbers.read*\end{tabular} &
  \begin{tabular}[c]{@{}l@{}}user\_birthday\\ user\_hometown\\ user\_gender\\ user\_age\_range\\ instagram\_graph\_user\_profile*\end{tabular} &
   &
  \begin{tabular}[c]{@{}l@{}}address\\ birthDate\\ phoneNumbers\\ backgroundPicture\end{tabular} \\ \hline
\textbf{Personal} &
  \begin{tabular}[c]{@{}l@{}}userinfo.profile\\ photoslibrary*\\ fitness*\\ tasks*\end{tabular} &
  \begin{tabular}[c]{@{}l@{}}user\_location\\ user\_photos\\ user\_videos\\ instagram\_graph\_user\_media*\end{tabular} &
   &
  geoLocation \\ \hline
\textbf{Behavioural} &
  \begin{tabular}[c]{@{}l@{}}games*\\ user.organization.read*\end{tabular} &
  \begin{tabular}[c]{@{}l@{}}user\_likes\\ user\_posts\\ user\_link\end{tabular} &
   &
  \begin{tabular}[c]{@{}l@{}}organizations\\ positions\\ educations\\ projects\\ certifications\\ skills\\ volunteeringInterests\\ volunteeringExperiences\end{tabular} \\ \hline
\textbf{Sensitive} &
  \begin{tabular}[c]{@{}l@{}}contacts\\ drive\\ gmail\\ documents*\\ spreadsheets*\\ youtube*\end{tabular} &
  user\_friends &
   &
  \begin{tabular}[c]{@{}l@{}}websites\\ industryName\\ courses\\ testScores\\ summary\end{tabular} \\ \hline
\end{tabular}
\label{tableProvidersComparison}
\end{table*}

\section{API Analysis of Identity Providers}
\label{sec.api.analysys}
User data available to third-party services through OAuth-backed APIs vary with each identity provider. We first compare IdPs before evaluating the RPs that request permissions via these APIs. The diverse native services provided by IdPs lead to differences in the types of user data available within each IdP's SSO platforms. These differences make it difficult to objectively compare permissions across multiple IdPs at different granularity. To assist with this comparison, we scanned OAuth-backed data attributes relating to personal user data for each IdP and categorised the attributes based on how the information could be used (or misused). We grouped items into five data categories: \textit{basic}, \textit{identity}, \textit{personal}, \textit{behavioural}, and \textit{sensitive}, as shown in Table~\ref{tableProvidersComparison}.

\subsubsection*{\textbf{Categories}}
Data in the \textit{basic} category includes the attributes we consider least privacy-invasive (e.g., name, email, profile image) shared with RPs through OAuth APIs. Here, name refers to the user's full name or profile username, and in the case of Apple SSO, the user is allowed to choose a custom name to be shared with the specific RP being accessed. Data attributes that facilitate identification of specific users such as user's birthday, gender, mobile number and street address are grouped into the \textit{identity} category. These are security-critical user data that, in the hands of adversaries can be misused to impersonate users (e.g., obtain a new mobile SIM). The \textit{personal} category includes data that enable RPs to access the user's personal data including photos, videos and current location. While such data pertains to users, it could also involve other secondary individuals who are part of the data being shared (e.g., a friend in the user's videos). We include data attributes (e.g., social content liked by users) that could exhibit a user's personality/habits as \textit{behavioural} data. User's \textit{behavioural} data can possibly be used by online trackers and other scripts to study users. Finally, we include user data such as emails, contact lists and documents that could contain personal information under the \textit{sensitive} category.

We note that some user data can be associated with multiple categories. For example, a passport copy uploaded to cloud storage can be both \textit{sensitive}- and \textit{identity}-related. In such cases, we include the attribute in the category we deemed more relevant. Our goal is to provide an overview of the types of user data accessed by RPs, and not a mutually-exclusive categorisation of OAuth-backed APIs. Table~\ref{tableProvidersComparison} groups relevant attributes from each provider into the five categories. We selected an initial list of attributes relevant to user data by reviewing OAuth documentation pages available on each IdP platform. Then, we refined this list to include all attributes requested by the top 500 sites in each of the five countries. In Fig.~\ref{figLoginForms}, we include screenshots of UI shown to SSO users of the four IdP. We now discuss the four IdP in the context of data categories made available through their OAuth-backed APIs.

\subsection{\textbf{Google OAuth API}}
Google's \textit{OAuth 2.0 Scopes} document~\cite{googleOauth2Scopes} categorizes data attributes into APIs based on the service where each is normally used within Google's ecosystem. For example, the Gmail API groups attributes relevant to sending and receiving emails in a Gmail inbox. This is useful for third-party email clients (e.g., Mozilla Thunderbird) that externally manage users' emails. Personal user information such as birthday, gender, and contact information are exposed through the People API. Other Google APIs that we found to be commonly used by RPs include the Calendar API (used for managing Google calendars) and the Drive API (used for accessing the user's Google Drive storage). Although not relevant to our study, Google makes available several other attributes to RPs through OAuth.

In addition to attributes found in our dataset, we included attributes related to personal user data from all Google APIs listed in the Scope document~\cite{googleOauth2Scopes}. Google's OAuth 2.0 APIs classify a subset of their attributes as belonging to \textit{sensitive} (different from \textit{sensitive} category defined in our study) and \textit{restricted} scopes. Most RP applications requesting access to any of these attributes must go through a verification process reviewed by Google prior to use~\cite{googleOauthVerificationFaqs}.

Google's OAuth 2.0 platform includes two types of profile-related attributes. The \texttt{profile}~\cite{googleApiAuth} attribute under \textit{basic} category shown in Table~\ref{tableProvidersComparison} includes only the user's name, email and public profile image (and is included by default in response to requests), whereas \texttt{userinfo.profile} includes all publicly available information from the user's Google profile and must be explicitly requested. Outside of the SSO platform, profile information can be accessed through Google services such as Hangouts, Maps, YouTube and Play\footnote{\url{https://support.google.com/accounts/answer/6304920}}.

\subsubsection*{\textbf{Authentication}}
Google's OAuth 2.0 platform supports the OpenID Connect attribute, \texttt{openid}, which allows RPs to identify users as specified by the OpenID Connect 1.0 specification~\cite{openIdConnect1Spec}. Including the \texttt{openid} attribute in the authentication request allows the RP to obtain from Google an ID token containing fields that assert the user's identity (e.g., username, token's issuer, token expiry timestamp). The fields are encoded as key-value pairs in the data portion of a JWT (JSON Web Token)~\cite{jwtrfc} object with a digital signature signed by Google.

\subsubsection*{\textbf{Supported flows}}
Google's SSO platform uses both OAuth 2.0 and OIDC specifications to support standard \textit{flows}, including the implicit and authorization code flows. RPs specify the flow using the \texttt{response\_type} parameter, as discussed in Section~\ref{sec.framework}. Additionally, the value \texttt{permission} can be used for the parameter to specify use of implicit flow\footnote{\url{https://developers.google.com/identity/sign-in/web/reference}}.

\begin{figure}[t]
    \includegraphics[width=\columnwidth]{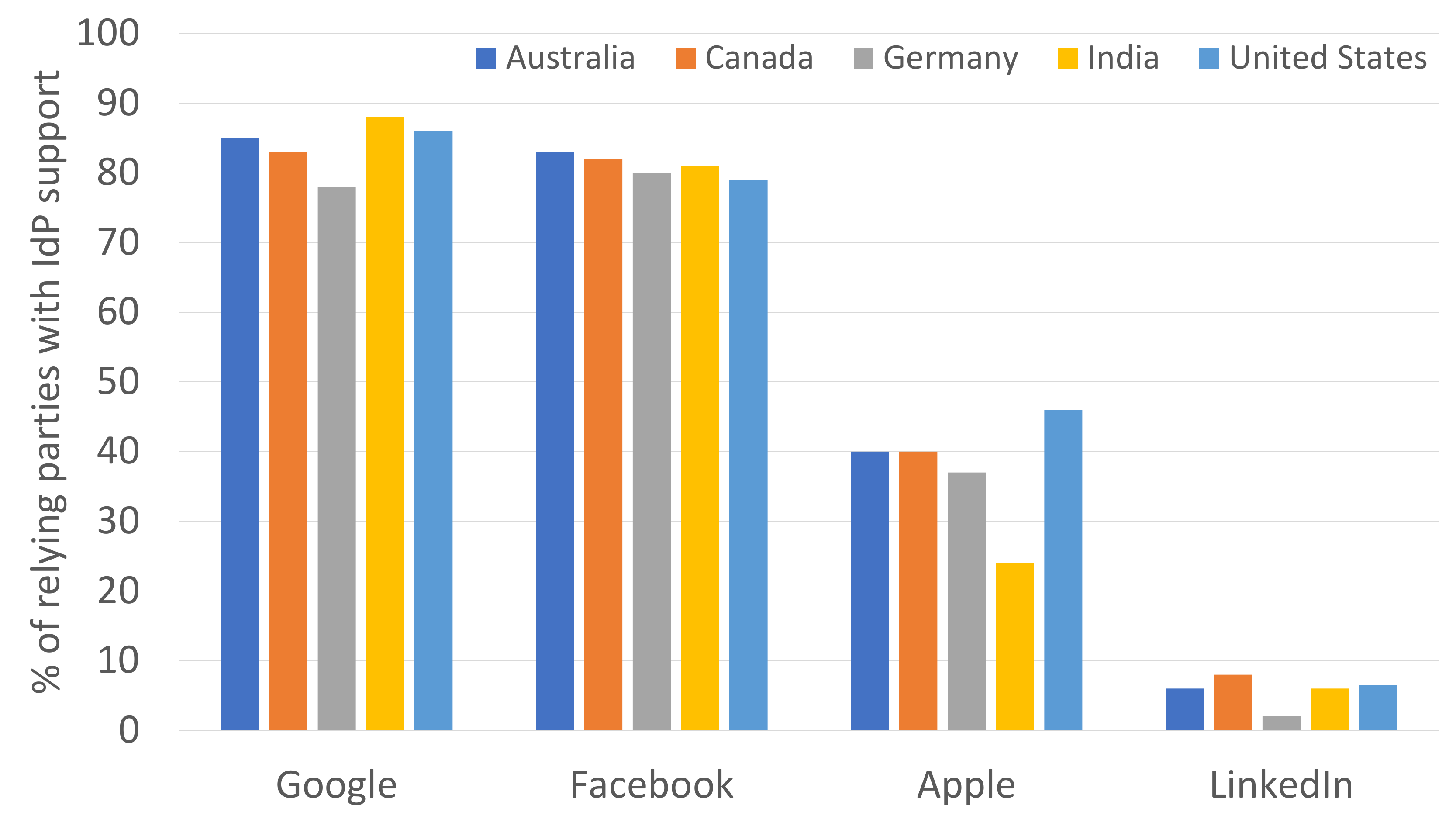}
    \caption{Percentage of RPs per country in our dataset that support each IdP.}
    \label{figProviders}
\end{figure}

\subsection{\textbf{Facebook OAuth API}} 
Facebook's Graph API~\cite{facebookGraphApi} is the platform for third-party applications to interact with a user's Facebook data. Applications implementing SSO with Facebook use the \textit{Facebook Login} interface to identify users. RPs can gain read-access to a wide range of data attributes from a user's Facebook profile. \textit{Facebook Login} provides social media-specific \textit{behavioural} data (e.g., Facebook pages likes and social posts created by the user) not available through other IdPs. We find that almost all of these attributes (as marked in Table~\ref{tableProvidersComparison}) are requested by a site in our empirical study.

Other permissions available in \textit{Facebook Login} allow applications to view, create, edit and delete content on user-administered Facebook Pages. This is also useful for business management applications to integrate business-owned Facebook Pages in promoting products and services. Facebook also provides Instagram Permissions for RPs (e.g., a service managing user's digital photo library) to access a user's Instagram content~\cite{facebookOauthPermissions}. Although these permissions are available for access, we focus our analysis on user-data attributes (listed on Table~\ref{tableProvidersComparison}) accessed by RPs in the top 500 sites.

All SSO requests (including those that do not explicitly request the \texttt{public\_profile} attribute) to \textit{Facebook Login} require the user to allow the RP access to default fields (Facebook name and profile picture)~\cite{facebookOauthPermissions}. RPs requesting permissions other than \texttt{public\_profile} (user's name and profile picture), \texttt{email} and \texttt{pages\_show\_list} (list of Facebook Pages managed by the user) are required to undergo a Facebook App Review~\cite{facebookAppReview}.

\subsubsection*{\textbf{Authentication}}
Although \textit{Facebook Login} does not support the OIDC specification, it allows RPs to authenticate users without requesting access to user data. Authentication can be done using \textit{Facebook Login}'s \textit{signed request} which is a signed base64url encoded JSON object containing a user ID issued by Facebook.

\begin{figure}[t]
    \includegraphics[width=\columnwidth]{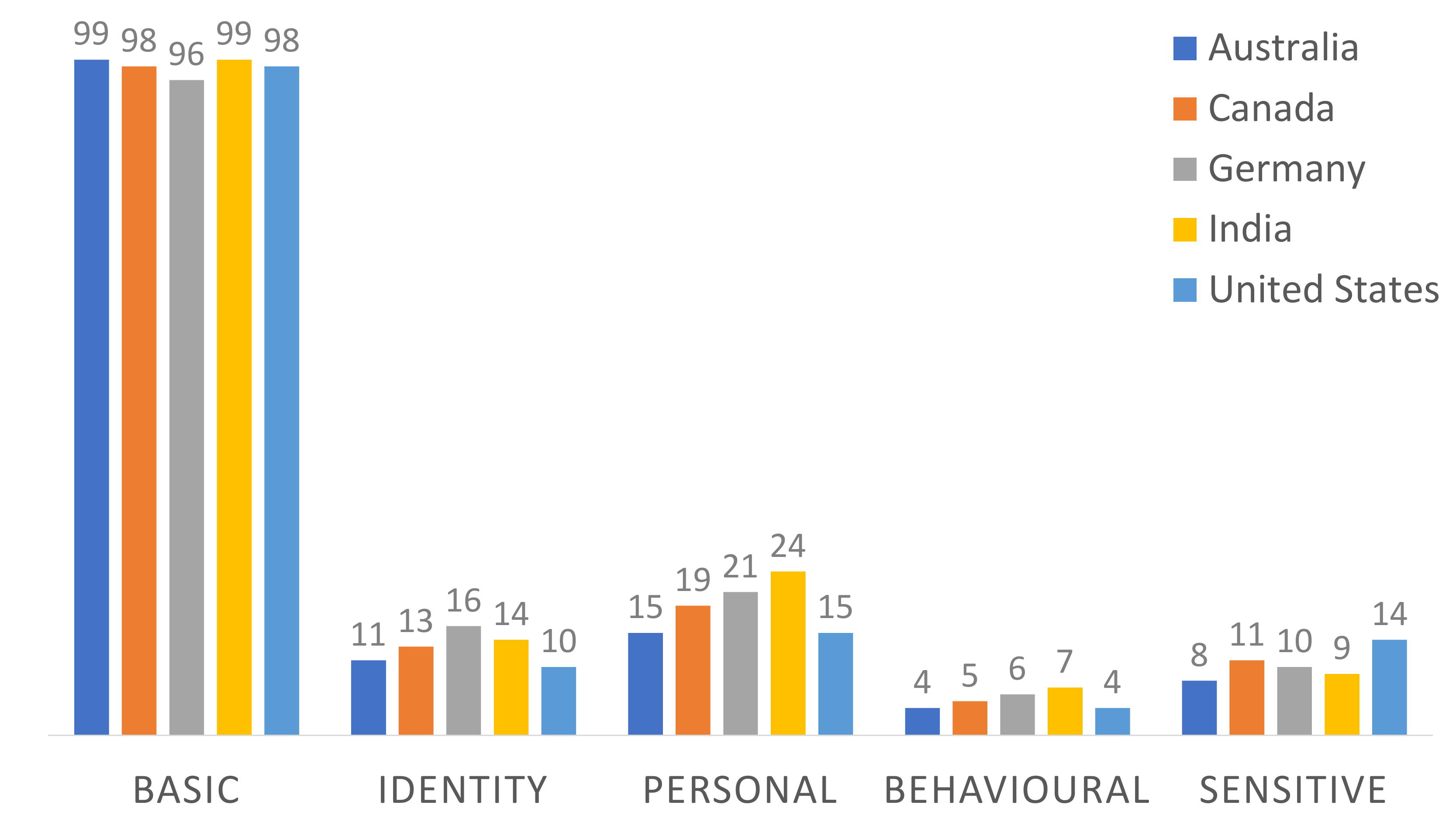}
    \caption{Percentage of RPs per country in our dataset requesting permissions per category (as defined in Table~\ref{tableProvidersComparison}). Some \textit{basic} attributes are required by default by IdPs, partially explaining high number of requests for this category.}
    \label{figDataAttributes}
\end{figure}

\subsubsection*{\textbf{Supported flows}}
\textit{Facebook Login} supports the standard OAuth 2.0 authorization code flow (\texttt{code}) and the implicit flow (\texttt{token}). RPs can additionally include the \texttt{signed\_request} parameter to obtain the user's ID and the \texttt{granted\_scopes} parameter to obtain a list of permissions approved by the user.

\subsection{\textbf{Apple OAuth API}}
\label{sec.signinwithapple}
Apple introduced the \textit{Sign in with Apple} framework in 2019 as a privacy-friendly alternative for SSO users wanting to use third-party web and mobile applications without disclosing their data. In addition to libraries for Apple devices, it includes a JavaScript library compatible with web applications in Windows or Android operating systems. \textit{Sign in with Apple} allows RPs to authenticate SSO users and to optionally request access to the user's \texttt{name} and \texttt{email} through the scope parameter. When requested, \textit{Sign in with Apple} allows users to either share their original email address or create an anonymous email address enabled by Apple's Private Email Relay Service. This service generates an anonymous email address unique to the user-RP pair and routes all email correspondence between RP and user through this email, hiding the user's real email from RP. \textit{Sign in with Apple} also helps RPs distinguish real users from bots through a boolean-value real user indicator~\cite{appleSignInDocumentation}. Applications on Apple's App Store using a third-party SSO service (e.g., Facebook, Google) are now required to also offer \textit{Sign in with Apple} as an SSO option\footnote{\url{https://developer.apple.com/news/?id=09122019b}}.

\subsubsection*{\textbf{Authentication}}
Unlike other IdPs, \textit{Sign in with Apple} is primarily used for authentication with limited user data (real or anonymous) available to RPs. Although the documentation~\cite{appleSignInDocumentation} lacks explicit mention of the OpenID standard, it closely follows OpenID conventions discussed in Section~\ref{sec.authentication}. A successful authentication returns an ID token in a JWT object, allowing the RP to identify the user.

\subsubsection*{\textbf{Supported flows}}
The accepted values for the \texttt{response\_type} parameter in \textit{Sign in with Apple} include either \texttt{code} (for authorization code flow), \texttt{id\_token} (for authentication) or both. It does not support the implicit flow.

\subsection{\textbf{LinkedIn OAuth API}}
LinkedIn is a widely used platform for sharing professional and personal user data. The \textit{Sign In with LinkedIn} platform allows users to authenticate and authorize profile access to third-party applications. In addition to \texttt{r\_emailaddress} (user's email address), \textit{Sign In with LinkedIn} provides three scope parameters related to a user's LinkedIn profile. Each parameter groups several attributes, providing less granular access to user data compared to other IdPs. Applications can specify the \texttt{r\_liteprofile} (or \texttt{r\_basicprofile} used in older version of the API) scope to request access to the user's full name, profile picture (including image meta data) and profile headline. The \texttt{r\_fullprofile} scope additionally includes all the other fields (listed in Table~\ref{tableProvidersComparison}) entered to the user's LinkedIn profile~\cite{signInWithLinkedIn}. For each SSO request, \textit{Sign In with LinkedIn} provides a unique user ID that allows RPs to authenticate users. If a request includes the \texttt{w\_member\_social} scope, the user can authorize the RP to create new LinkedIn posts on their behalf.

\definecolor{Gray}{gray}{0.9}

\begin{table*}[h]
\renewcommand*\arraystretch{1.4}
\centering
\footnotesize
\caption{OAuth 2.0 and related OpenID Connect (OIDC) flows used by RPs per country. N = total number of sites (among top 500) offering the IdP per country (number of RPs); n = number of RPs using the given flow; \% = percentage of RPs using a given flow (i.e., $\% = n/N*100$). Hybrid flows represent multi-valued response types described in Section~\ref{sec.authentication}.}
\label{tableFlows}
\begin{tabular}{l@{\hspace{1pt}}l@{\hspace{1pt}}l@{\hspace{1pt}}rrrrrrrrrrrrrrr}
\toprule
\textbf{IdP} &
  \multicolumn{1}{l}{\textbf{OAuth 2.0/OIDC Flow}} &
  \multicolumn{1}{l}{\textbf{Response Type}} &
  \multicolumn{3}{c}{\textbf{Australia}} &
  \multicolumn{3}{c}{\textbf{Canada}} &
  \multicolumn{3}{c}{\textbf{Germany}} &
  \multicolumn{3}{c}{\textbf{India}} &
  \multicolumn{3}{c}{\textbf{US}} \\ 

 &  &  & \textbf{N} & \textbf{n} & \cellcolor{Gray}\textbf{\%}  
 & \textbf{N} & \textbf{n} & \cellcolor{Gray}\textbf{\%}  
 & \textbf{N} & \textbf{n} & \cellcolor{Gray}\textbf{\%}    
 & \textbf{N} & \textbf{n} & \cellcolor{Gray}\textbf{\%}  
 & \textbf{N} & \textbf{n} & \cellcolor{Gray}\textbf{\%}   \\ \toprule

%%%%%%%%%%GOOGLE
\multirow[t]{8}{*}{Google} & 
  Authorization code & code & 148 & 95 & \cellcolor{Gray}64  & 132 & 85 & \cellcolor{Gray}64  & 98 & 74 &\cellcolor{Gray}76  & 151 & 83 & \cellcolor{Gray}55  & 159 & 95 &\cellcolor{Gray}60  \\ \cmidrule{2-18} 
 &
 Implicit &
  \begin{tabular}[t]{@{}l@{}}token\\ id\_token\\ id\_token token\end{tabular} &

  \begin{tabular}[t]{@{}r@{}}$\uparrow$\\ \\ \end{tabular} &
  \begin{tabular}[t]{@{}r@{}}4\\ 2\\ 34\end{tabular} &
  \cellcolor{Gray}\begin{tabular}[t]{@{}r@{}}3\\ 1\\ 23\end{tabular} &
  
    \begin{tabular}[t]{@{}r@{}}$\uparrow$\\ \\ \end{tabular} &
    \begin{tabular}[t]{@{}r@{}}4\\ 1\\ 35\end{tabular} &
    \cellcolor{Gray} \begin{tabular}[t]{@{}r@{}}3\\ 1\\ 27\end{tabular} &
 
     \begin{tabular}[t]{@{}r@{}}$\uparrow$\\ \\ \end{tabular} &
    \begin{tabular}[t]{@{}r@{}}0\\ 0\\ 17\end{tabular} &
    \cellcolor{Gray} \begin{tabular}[t]{@{}r@{}}-\\ -\\ 17\end{tabular} &
 
   \begin{tabular}[t]{@{}r@{}}$\uparrow$\\ \\ \end{tabular} &
  \begin{tabular}[t]{@{}r@{}}3\\ 0\\ 55\end{tabular} &
  \cellcolor{Gray}\begin{tabular}[t]{@{}r@{}}2\\ -\\ 36\end{tabular} &

 \begin{tabular}[t]{@{}r@{}}$\uparrow$\\ \\ \end{tabular} &
  \begin{tabular}[t]{@{}r@{}}2\\ 2\\ 50\end{tabular} &
\cellcolor{Gray} \begin{tabular}[t]{@{}r@{}}1\\ 1\\ 31\end{tabular} \\ \cmidrule{2-18}

 &
  Hybrid &
  \begin{tabular}[t]{@{}l@{}}code id\_token\\ code token\\ code id\_token token\end{tabular} &
  \begin{tabular}[t]{@{}r@{}}\\ \\ \end{tabular} &
  \begin{tabular}[t]{@{}r@{}}0\\ 0\\ 13\end{tabular} &
  \cellcolor{Gray}\begin{tabular}[t]{@{}r@{}}-\\ -\\ 9\end{tabular} &

    \begin{tabular}[t]{@{}r@{}}\\ \\ \end{tabular} &
    \begin{tabular}[t]{@{}r@{}}0\\ 0\\ 7\end{tabular} &
  \cellcolor{Gray}\begin{tabular}[t]{@{}r@{}}-\\ -\\ 5\end{tabular} &
  
     \begin{tabular}[t]{@{}r@{}}\\ \\ \end{tabular} &
    \begin{tabular}[t]{@{}r@{}}0\\ 0\\ 7\end{tabular} &
  \cellcolor{Gray}\begin{tabular}[t]{@{}r@{}}-\\ -\\ 7\end{tabular} &
  
     \begin{tabular}[t]{@{}r@{}}\\ \\ \end{tabular} &
    \begin{tabular}[t]{@{}r@{}}0\\ 0\\ 10\end{tabular} &
  \cellcolor{Gray} \begin{tabular}[t]{@{}r@{}}-\\ -\\ 7\end{tabular} &
  
     \begin{tabular}[t]{@{}r@{}}\\ \\ \end{tabular} &
    \begin{tabular}[t]{@{}r@{}}0\\ 1\\ 9\end{tabular} &
  \cellcolor{Gray} \begin{tabular}[t]{@{}r@{}}-\\ 1\\ 6\end{tabular} \\ \midrule

%%%%%%%%%%%FACEBOOK
\multirow[t]{3}{*}{Facebook} & Authorization code 
& code & 147 & 102 & \cellcolor{Gray}69  
&  134 & 90 & \cellcolor{Gray} 67  
& 103 & 73 &\cellcolor{Gray}71  
& 144 & 82 & \cellcolor{Gray}57  
& 148 & 84 & \cellcolor{Gray}57  \\ \cmidrule{2-18} 

 &
  Implicit &
  \begin{tabular}[t]{@{}l@{}}token\\ token signed\_request\end{tabular} &

  \begin{tabular}[t]{@{}r@{}}$\uparrow$\\ \end{tabular} &
  \begin{tabular}[t]{@{}r@{}}4\\ 41\end{tabular} &
 \cellcolor{Gray} \begin{tabular}[t]{@{}r@{}}3\\ 28\end{tabular} &
 
     \begin{tabular}[t]{@{}r@{}}$\uparrow$\\ \end{tabular} &
    \begin{tabular}[t]{@{}r@{}}5\\ 39\end{tabular} &
  \cellcolor{Gray} \begin{tabular}[t]{@{}r@{}}4\\ 29\end{tabular} &
  
       \begin{tabular}[t]{@{}r@{}}$\uparrow$\\ \end{tabular} &
    \begin{tabular}[t]{@{}r@{}}0\\ 30\end{tabular} &
 \cellcolor{Gray} \begin{tabular}[t]{@{}r@{}}-\\ 29\end{tabular} &
 
      \begin{tabular}[t]{@{}r@{}}$\uparrow$\\ \end{tabular} &
    \begin{tabular}[t]{@{}r@{}}2\\ 60\end{tabular} &
  \cellcolor{Gray}\begin{tabular}[t]{@{}r@{}}1\\ 42\end{tabular} &
 
      \begin{tabular}[t]{@{}r@{}}$\uparrow$\\ \end{tabular} &
    \begin{tabular}[t]{@{}r@{}}3\\ 61\end{tabular} &
 \cellcolor{Gray} \begin{tabular}[t]{@{}r@{}}2\\ 41\end{tabular} \\ \midrule

%%%%%%%%%%APPLE
\multirow[t]{2}{*}{Apple} & Authorization code & code 
& 70 & 34 & \cellcolor{Gray} 49 
& 64 & 30 & \cellcolor{Gray} 47 
&47 & 30 & \cellcolor{Gray}64 
& 42 & 16 &  \cellcolor{Gray}38 
& 85 & 35 & \cellcolor{Gray}41 \\
\cmidrule{2-18}

&
  Hybrid & code id\_token 
  & $\uparrow$ & 36 & \cellcolor{Gray}51 
  & $\uparrow$ & 34 & \cellcolor{Gray}53 
  & $\uparrow$ & 17 &\cellcolor{Gray}36 
  & $\uparrow$ & 26 & \cellcolor{Gray}62 
  & $\uparrow$ & 50 & \cellcolor{Gray}59\\ \midrule
 
%%%%%%%%LINKEDIN 
LinkedIn & Authorization code & code 
& 9 & 9 & \cellcolor{Gray}100 
& 11 & 11 & \cellcolor{Gray}100 
& 3 & 3 & \cellcolor{Gray}100 
& 10 & 10 & \cellcolor{Gray}100 
& 11 & 11 & \cellcolor{Gray}100 \\ \bottomrule
\end{tabular}
\end{table*}

\begin{figure}
    \includegraphics[width=\columnwidth]{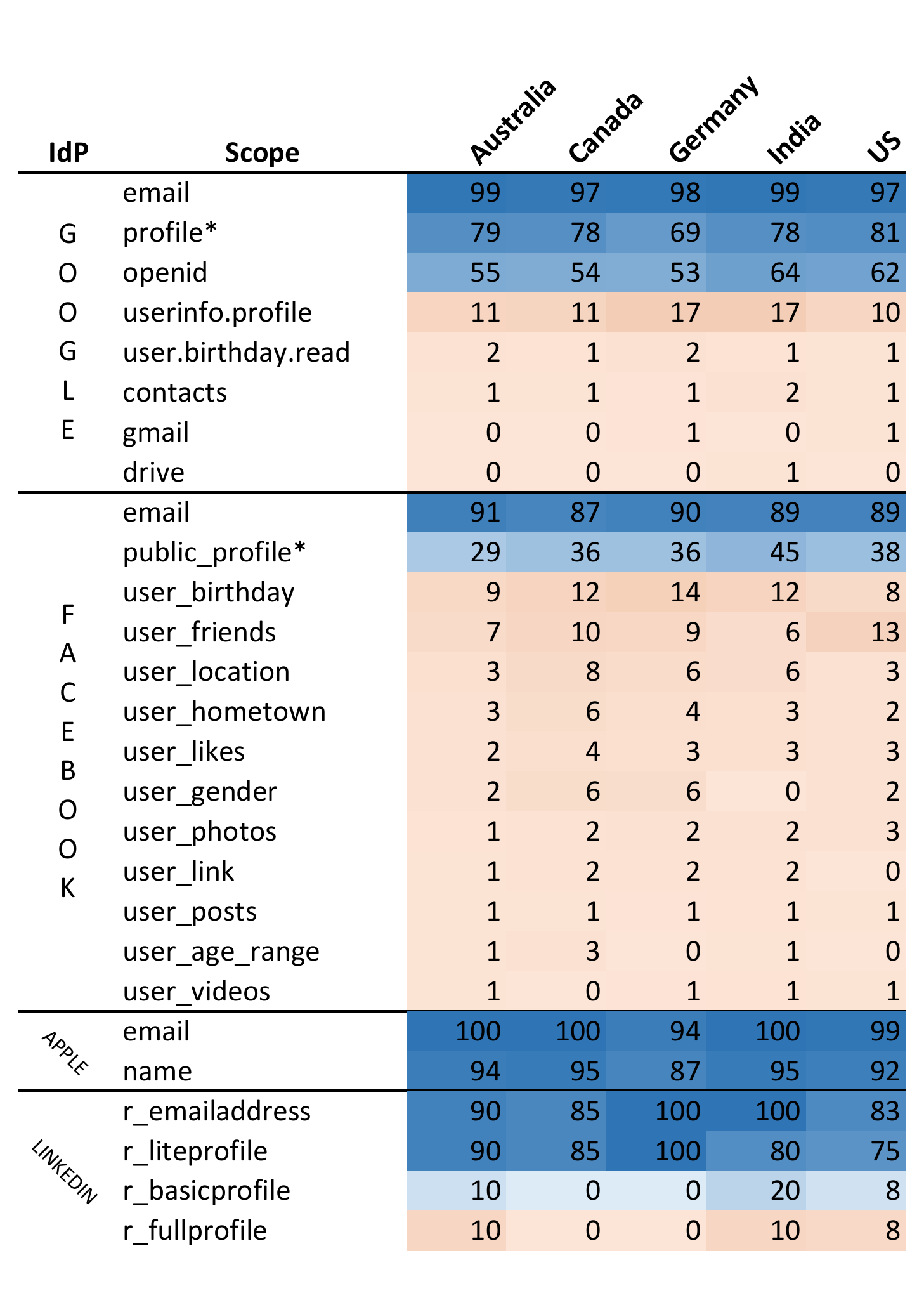}
    \caption{The percentage of RPs per IdP in our dataset that requests a particular scope attribute. Blue cells represent attributes from the \textit{basic} category. Darker cells indicate a higher percentage of RPs making a given request. * identifies default scope attributes that the IdP requires users to share with RP.}
    \label{figHeatmap}
\end{figure}

\subsubsection*{\textbf{Authentication}}
In LinkedIn's documentation for \textit{Sign In with LinkedIn}, we did not find explicit mention of the OpenID standard or information on whether relying parties could request only the member ID, a unique identifier specific to the RP-user pair. However, RPs can authenticate users using the OAuth flows supported by the platform. After a successful login, the RP obtains an OAuth 2.0 access token which includes the user's LinkedIn member ID.

\subsubsection*{\textbf{Supported flows}}
\textit{Sign In with LinkedIn} defines two types of \textit{consents} (analogous to OAuth flows): \textit{member authorization} and \textit{application authorization}. We concern ourselves only with \textit{member authorization}, where \textit{Sign In with LinkedIn} requires the use of the authorization code flow from OAuth 2.0. Conversely, \textit{application authorization} uses OAuth 2.0's \textit{client credentials flow} for systems requiring machine-to-machine authorization, without user involvement~\cite{oauth2rfc}, and is outside the scope of this study.

\section{Empirical Results}
\label{sec.results}
In this section, we report the findings of our empirical study on the use of OAuth 2.0 to access user data in popular SSO services.

\subsection{Distribution of Providers}
Our results show that Google and Facebook are consistently the most popular SSO options in top websites across all five countries. Apple is currently the third most popular option, possibly due to its relatively recent introduction of \textit{Sign in with Apple} in 2019. As shown in Fig.~\ref{figProviders}, \textit{Sign in with Apple} is less popular in India, which is consistent with Apple's lack of popularity in India~\cite{cnbcApplePopularityInIndia}. However, recent requirements (discussed in~\ref{sec.signinwithapple}) for apps on Apple's App Store could lead to an increase in its use with RPs.

\begin{figure*}[]
    \includegraphics[height=.94\textheight]{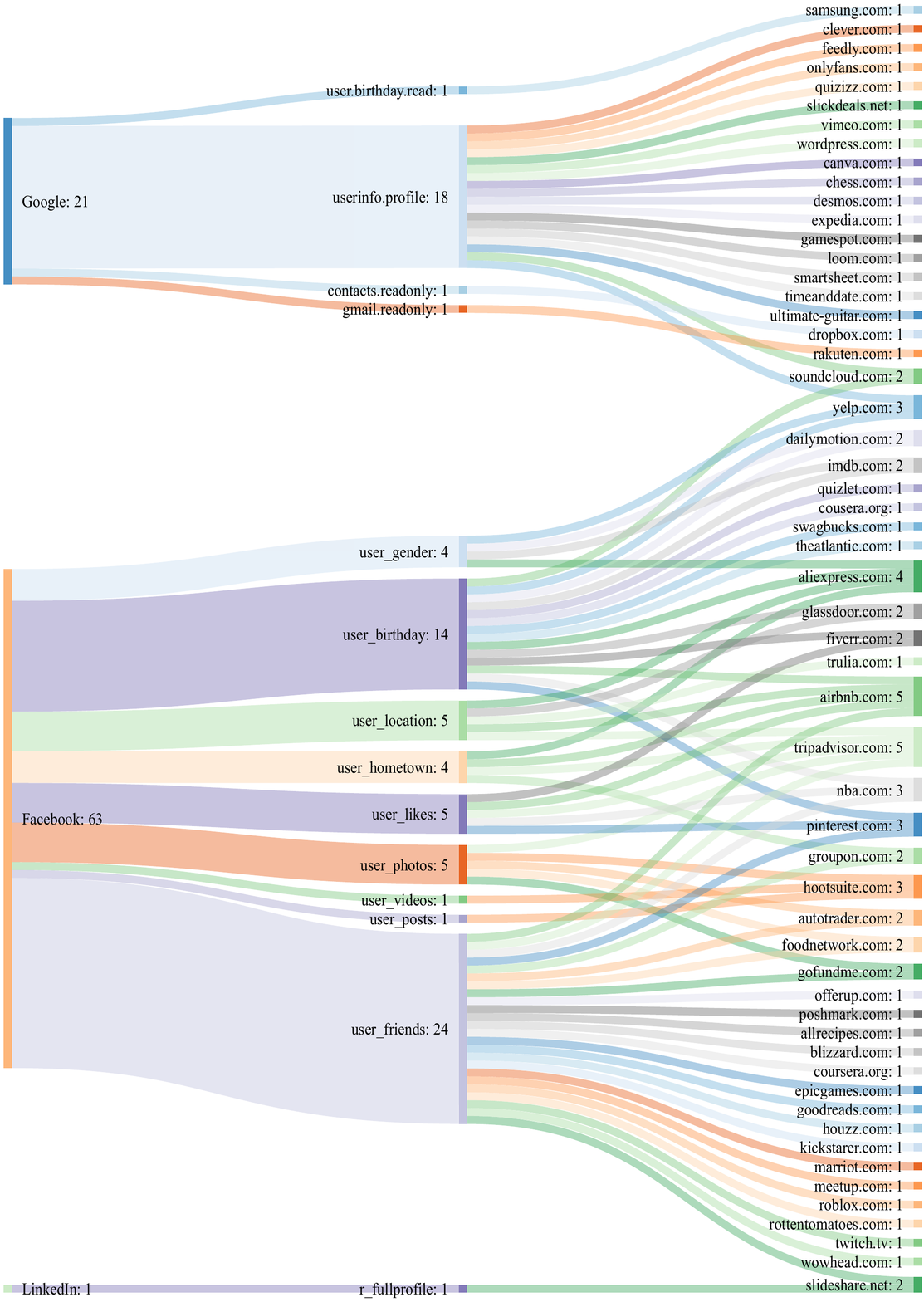}
    \caption{Data attributes (from Table~\ref{tableProvidersComparison}) requested by relying parties in the top 500 US sites. For readability, this diagram excludes the \textit{basic} category. Apple is not included in this chart since it only supports \textit{basic} attributes.}
    \label{figUSSankey}
\end{figure*}

\begin{figure}[]
    \includegraphics[width=\columnwidth]{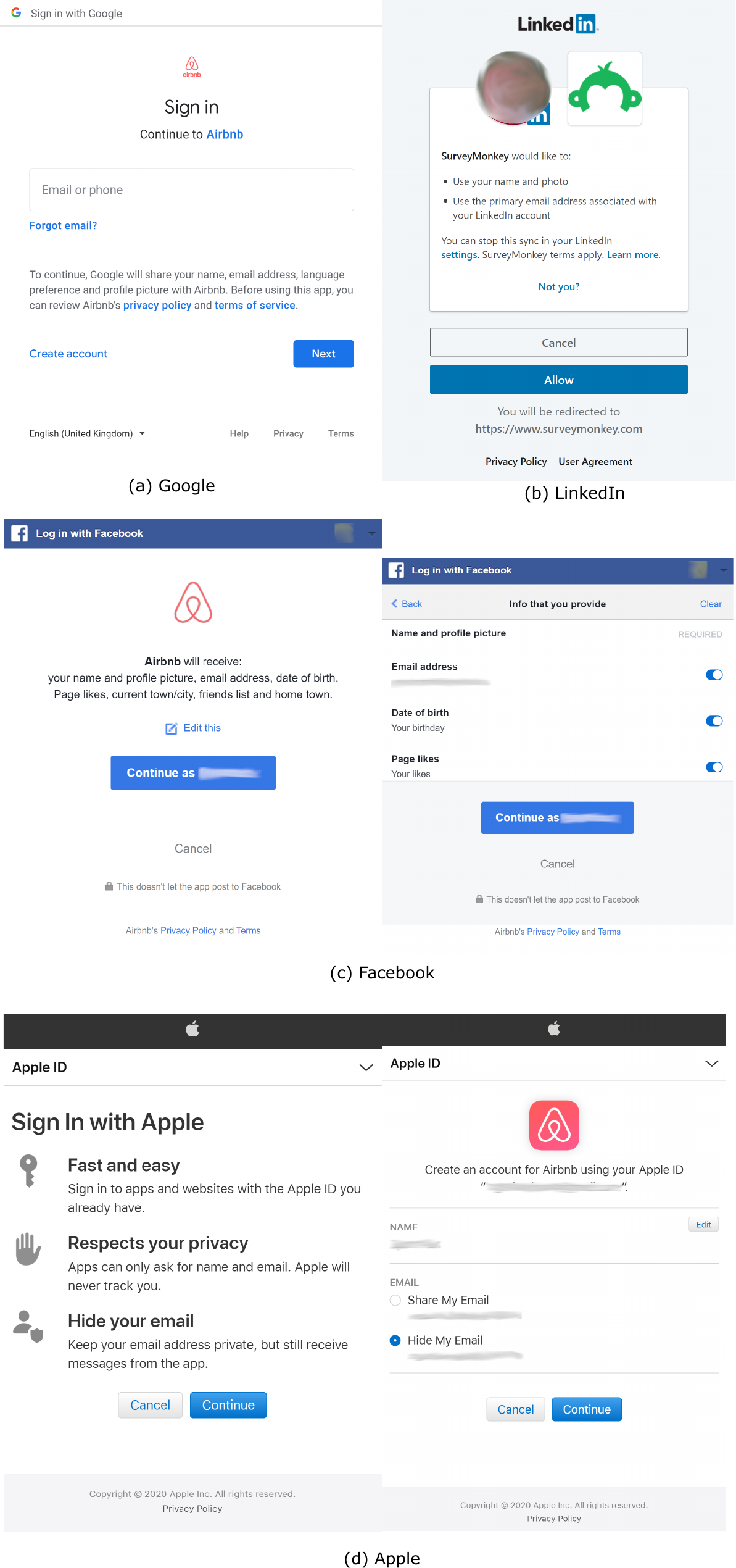}
    \caption{UI of SSO login forms on IdP sites. (a) Google and (b) LinkedIn do not allow users to alter fine-grained permissions granted to RPs. (c) Facebook allows users to selectively opt-out of non-\textit{default} permissions requested by an RP. (d) Apple allows users to use a substitute name and anonymous email with an RP. In cases of (b), (c) and (d), the IdP presents its default login dialog before showing the SSO screens if the user is not already logged in, as discussed inline. Personal details are greyed out in these images.}
    \label{figLoginForms}
\end{figure}

\subsection{Comparing requested data across countries}
\label{sec.results.dataRequestsAcrossCountries}
Fig.~\ref{figHeatmap} provides a sorted list of the most requested attributes for each IdP. The majority of the RPs request one or more \textit{basic} attributes that help identify users (e.g., to display the user's name and profile picture on the RP site). Although relatively similar patterns emerge across countries, we do note some variation on particular attributes. For example, Google's \texttt{userinfo.profile} is requested more frequently in Germany and India, while LinkedIn's \texttt{r\_basicprofile} and \texttt{r\_fullprofile} are requested most frequently in India, followed by Australia, and not requested at all in Canada and Germany. These variations could be a result of an IdP's popularity with the users in a given country. Perceived sensitivity of specific user data could also vary across countries, and consequently lead to different privacy implications for users. 

During our analysis, we also observed that RPs use different web designs and make available different SSO options in different countries. For example, Rakuten.com (a popular e-commerce marketplace) supports three SSO options (Google, Facebook and Apple) for US users and requests read access to the user's emails when signed on with Google SSO. However, Rakuten.ca (for Canadian users) only supports Facebook and Apple SSO options. Interestingly, Rakuten.de (for German users) does not support any SSO options. We did not have any Rakuten sites specific to users in Australia or India in our dataset.

\subsection{Comparing requested data across providers}
\label{sec.results.dataRequestsAcrossProviders}
Although each IdP exposes different user data, we notice differences in which attributes RPs requested for similar data categories across the providers. For example, $13\%$ of RPs in the US request access to users' friends list with the Facebook SSO, but only $1\%$ request the contacts list with the Google SSO. Fig.~\ref{figUSSankey} links RPs in the US sites with the user data they request from each IdP. For readability, this chart does not include the \textit{basic} attributes, and instead focuses on RPs that request more sensitive non-\textit{basic} attributes. From Fig.~\ref{figUSSankey}, we see that the US airbnb.com and tripadvisor.com request more non-\textit{basic} attributes than other RPs (towards the center of the graph, each with 5 attributes from Facebook).

Both Facebook and Google SSOs are supported by tripadvisor.com. When a user logs in with Facebook, tripadvisor.com requests access to user's hometown, location (as listed in profile), list of likes (of Facebook pages), friends list and photos. However, the request only includes the \textit{basic} attributes (\texttt{email}, \texttt{openid}, and \texttt{profile}) with Google SSO, and thus these do not appear in Fig.~\ref{figUSSankey}.  

\begin{figure*}[]
    \centering
    \includegraphics[width=.85\textwidth]{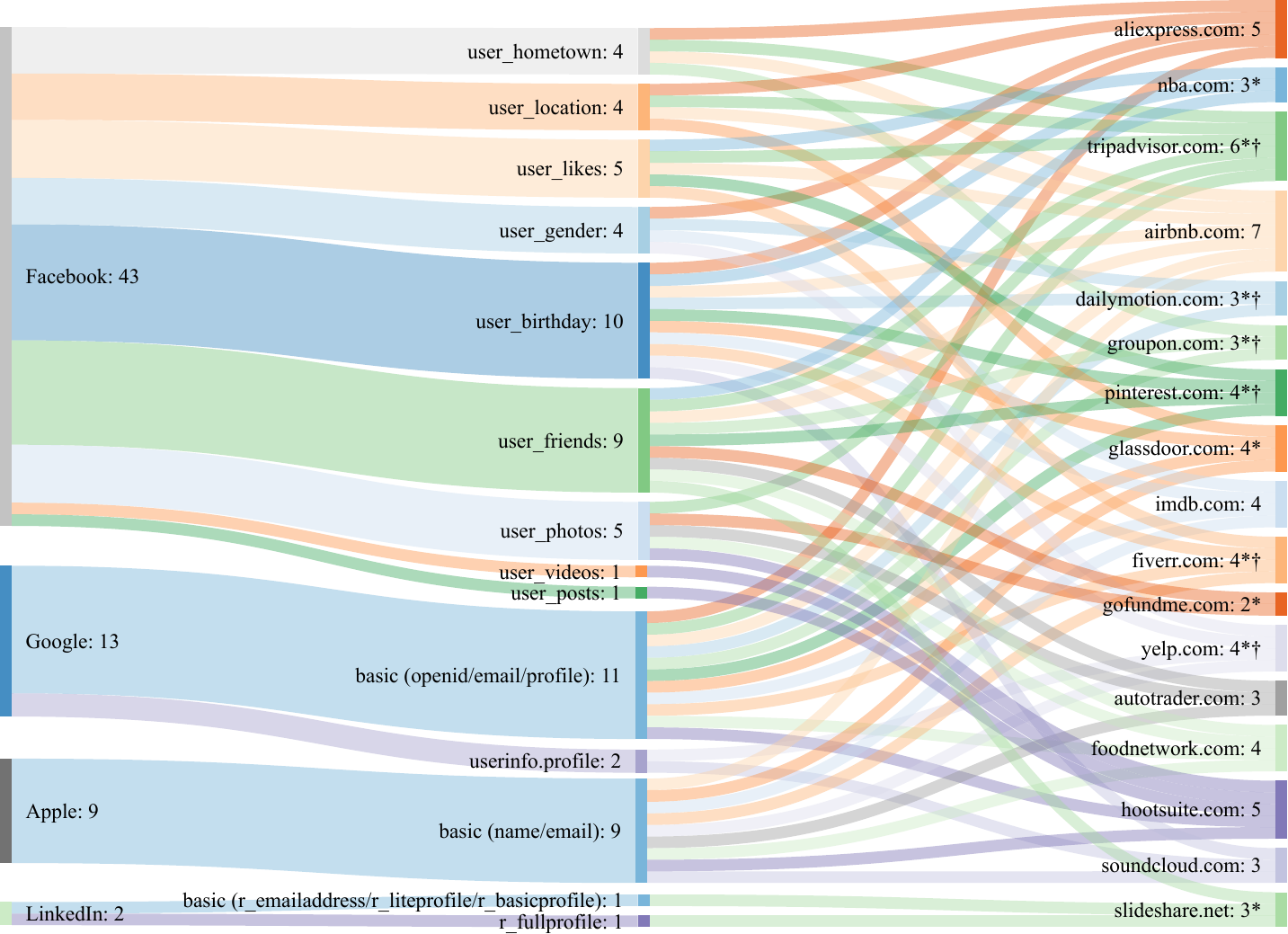}
    \caption{Comparison of all data attributes requested (including from the \textit{basic} category) by a subset of US RPs. The included RPs request at least 2 non-\textit{basic} permissions. For readability, we exclude redundant connections to the \textit{basic} category for RPs that already request one or more non-\textit{basic} attributes with an IdP, since such requests include \textit{default} fields. RPs using client-side flows in OAuth 2.0 are shown with *(Facebook) and $\dagger$(Google). The main takeaways are explained inline.}
    \label{figUSSankeySubset}
\end{figure*}

To offer a more complete example, we provide Fig.~\ref{figUSSankeySubset}. Here, we identified a subset of RPs that request $2$ or more non-\textit{basic} attributes (from Fig.~\ref{figUSSankey}) and also include the \textit{basic} attributes requested from other IdPs. Out of the $17$ RPs shown in Fig.~\ref{figUSSankeySubset}, $14$ support an alternate SSO login requesting only the \textit{basic} attributes, suggesting that, in these cases, some user options are less privacy-invasive than others. For the remaining RPs in the figure, nba.com and gofundme.com support only Facebook as a SSO option; slideshare.net supports two SSOs but neither option uses only \textit{basic} attributes.

\subsection{Use of OAuth 2.0 and related OIDC Flows}
\label{sec.results.flows}
Table~\ref{tableFlows} summarizes the different OAuth 2.0/OpenID Connect flows used by RPs in our dataset. Of particular concern is that a significant number of RPs use the less secure implicit flow, especially in India ($38\%$ with Google SSO, $43\%$ with Facebook SSO) and US ($33\%$ with Google SSO and $43\%$ with Facebook SSO). RPs using implicit flow receive access tokens in the user's browser directly from the token endpoint. Apple and LinkedIn do not support the implicit flows on their SSO platforms, thus forcing all RPs to use more secure options. As shown in Table~\ref{tableFlows}, both Google and Facebook SSOs support both the client-side and server-side flows.  From our analysis of countries, we also note that fewer RPs in Germany use the less secure implicit flows compared to other countries.

\section{Privacy Implications}
\label{sec.privacy.implications}
In this section, we discuss privacy implications from our evaluation of OAuth 2.0 use in popular SSO services.

\subsection{Impact of interface design}
Fig.~\ref{figLoginForms} shows the design of SSO user interfaces (UI) presented to the user when they choose to login using SSO.
Usable SSO interfaces are essential in informing users about the permissions requested by an RP. Designs should be as simple as possible and provide users with a path-of-least-resistance to securely complete authentication tasks~\cite[\S9.8]{van2020computer}. IdPs must obtain informed consent during SSO workflows, which theoretically provides users with some control over their privacy. However, earlier research on Android's permission system, which tackles a similar issue, has shown that permission warnings are ineffective in informing users about the risks associated with allowing access to applications~\cite{felt2012android}. Android permission warnings focus on resource access but lack useful information for users to understand the associated benefits and risks~\cite{reardon201950}. Even when users read permission warnings, they are unaware of risks and simply trust the marketplaces to have reviewed the hosted applications~\cite{kelley2012conundrum}. We draw parallels to SSO permission requests and, through our inspection, observe that the existing SSO UI designs similarly lack useful information for users to convey risks associated with sharing personal data. They also do not convey the value provided to the user from granting RPs access to their data.

Information about requested permissions is presented differently by each IdP (see Fig.~\ref{figLoginForms}). If the user is not already logged in, some IdPs (i.e., LinkedIn, Facebook and Apple) present their default login screens before showing SSO screens. This means that redirected users can view the requested permissions only after they complete authentication with the IdP. This may lead to users granting permissions they would have chosen to deny if the information was presented before they logged in. While some IdPs allow users to edit the requested permissions (shown in Fig.~\ref{figLoginForms}), others require users to grant all the requested permissions. Given the differences on the amount of private data requested among available SSO options, we highlight the importance of presenting the requested permissions to users prior to their decision to login with a specific provider.

\subsection{Implications due to Implicit Flow}
The OAuth 2.0 implicit flow was created due to past browser restrictions limiting websites to making requests only within its own domain~\cite{oktaImplicitDead}. This prevented JavaScript-based apps from using the authorization code flow since it involves making requests to the IdP domain, which is always different when the RP and IdP are different entities. As discussed in Section~\ref{sec.framework.implicit}, modern browsers support cross-origin requests, allowing the use of more secure flows. Our analysis shows that many RPs still use the implicit flow that returns access tokens in the redirection URL. This is a security concern since URLs are persisted in users' browsing history and it increases the attack surface for access token leakage. Access tokens are a type of ``bearer'' tokens, and any party in possession can use it to access protected resources without involving the user or RP. Access tokens provide access to specific resources for a limited duration. Users may not be aware of the risks associated with access token leaks, especially when the RP uses client-side flows. The potential damage to users increases when the access token's scope allows excessive access to sensitive user data. RPs can reduce the attack surface by requesting minimum access and using secure flows.

\subsection{Offline data leaks}
RPs and IdPs should clearly indicate to users the purpose of requested permissions prior to issuing access tokens.
OAuth enables an RP to improve usability for users through customization based on user-data attributes from an IdP. Once granted access, the RP is able to download and persist user data for further processing. As mentioned in Section~\ref{sec.api.analysys}, many IdPs review RP applications that request access to sensitive user data. IdPs also provide an interface for users to revoke previously granted access to an RP, invalidating all access tokens issued to the RP. However, this does not prevent a rogue RP from misusing any user data already accessed. Since user data is processed on RP applications (not controlled by IdP), it is not possible for the user or IdP to be aware of any misuse of accessed data.

Without proper security measures, even a well-intentioned RP can be vulnerable to data breaches that increase the attack surface for users and their data. Although using OAuth ensures that the user's passwords are safe from an attack on RP, leaked access tokens can equally cause damage by allowing attackers to access user data~\cite{HackGithubTokens}. It can be challenging for users to track attacks on RPs, and understanding the implications requires a mental model of the SSO system that many users lack~\cite{sun2011makes}. Another challenge for users involves access to decommissioned accounts. RPs identify users by trusting the IdP's verification of user credentials. If a user has stopped using an RP or de-linked their IdP account with RP, it may not be possible for the user to later correspond with the RP (e.g., to demand deletion of personal data).

\section{Toward privacy-friendly OAuth SSO}
\label{sec.recommendations}
Having uncovered examples of services offering a variety
of SSO IdP alternatives but with
major differences---largely invisible to users---in the 
categories and amounts of personal user data accessed,
we now ask: How might \textit{privacy},
and the \textit{transparency} and \textit{accountability} of RPs,
be improved?
To this end, we suggest changes below---as much to
encourage discussion and exploration of
such protocol and interface changes, as for  
any merit in the specific suggestions per se.
Our discussion briefly notes
the impact on and roles of four stakeholders:
Users, IdPs, RPs, and OAuth Specification authors.

\begin{itemize}
\item [C1:] 
\textit{When registering with an IdP,
an RP could be required to provide descriptions justifying  
each OAuth user-data attribute they plan to request from users.}
\end{itemize}

\noindent
This might introduce \textit{value labels} (cf.\ \textit{privacy nutrition labels}\footnote{\url{https://cups.cs.cmu.edu/privacyLabel/}})
and convey intended uses and benefits (if any) to users.  

\begin{itemize}
\item [C2:] \textit{Each RP could disclose to the user before they select an IdP whether one IdP choice will access more user data than others and provide an explanation of potential user benefits.}
\end{itemize}

\begin{itemize}
\item [C3:] 
\textit{The RP could display information justifying each OAuth user-data attribute requested, before asking a user to select an IdP.} 
\end{itemize}

\noindent
Such information might be conveyed via a second-level RP user interface.
Recall that OAuth next redirects users to the IdP.

\begin{itemize}
\item [C4:] \textit{The IdP could re-iterate the justification (e.g., value labels) before asking users
to authorize release of RP-requested data attributes.}  

\item [C5:] \textit{IdPs could enforce exclusive use of server-side flows (i.e., disallow implicit flow) for any RP request involving access to sensitive user data.} 
\end{itemize}

\noindent
Here ``sensitive'' is as defined in Table~\ref{tableProvidersComparison}.
Our motivation is that widely scoped access tokens create greater risks,
and client-side flows increase the attack surface (see Section~\ref{sec.framework.implicit}).
The OAuth spec could mandate the above, as well as the following.

\begin{itemize}
\item [C6:] \textit{The OAuth spec could allow
optional scope parameters distinct from those denoted mandatory for RP operation.}  

\end{itemize}

\noindent
Rather than the present all-or-none scenario (forcing users to accept all, or abandon the IdP for a given RP visit),
privacy-conscious users could opt-out of selected \textit{optional} parameters.

The above changes could allow audits or privacy compliance checks (by IdP or third parties),
and support informed choices by privacy-conscious users.
Over time, this could result in RP-IdP pairs 
following the privacy best practice of requesting only \textit{need-to-know} data,
and privacy-friendly IdPs gaining a ``preferred-IdP'' status based on their
user interface quality, and auditing of RP compliance with new IdP-to-RP guidelines for how RPs should display \textit{value labels} to users.
We suggest two further items that may help privacy-aware users make informed decisions.

\begin{itemize}
\item [C7:] \textit{A third-party tool could be built
to shed light on the consequences of different IdP (OAuth attribute) choices.}\footnote{Perhaps analogous to AppCensus  
for mobile app permissions (\cite{appcensus}; Reardon et al.~\cite{reardon201950}).}  

\item [C8:] \textit{Reputation-based or data-driven community efforts could provide
privacy ratings for SSO options at popular RPs.}
\noindent
\end{itemize}

\noindent
Many of these suggestions may of course face numerous hurdles,
one being that the agendas of commercial RPs and IdPs  
are often not aligned with those of privacy-conscious users.  
However, we argue that awareness and
discussion of technical possibilities are important steps towards supporting user privacy, and shedding light on any RP \textit{dark patterns}~\cite{narayanan2020dark}.

\section{Related Work}
\label{sec.related.work}
Zhou et al.~\cite{zhou2014ssoscan} built SSOScan, a black-box testing tool to automatically scan the top 20,000 US sites and found at least one serious vulnerability due to implementation flaws in 345 of the 1660 sites that supported Facebook SSO in 2014. Drakonakis et al.~\cite{drakonakis2020cookie} built an auditing framework for evaluating web applications for implementation flaws related to authentication and authorization, including applications that support SSO logins. They simulate user interactions to automatically create accounts and login to 25K websites to find that 9,324 domains are vulnerable to leaking sensitive user data to unauthorized parties. Their approach to simulate user interactions and automatically obtain SSO protocol-related information is similar to ours but as noted in our introduction, instead of privacy leaks to unauthorized parties, we evaluate privacy implications in websites that are explicitly granted (although users might not be aware) access to the user's personal data protected by SSO providers. Mainka et al.~\cite{mainka2016do} design a testing framework to investigate malicious IdPs in OpenID implementations and identify four novel attack classes affecting 11 of 16 systems tested. Fett et al.~\cite{fett2016comprehensive} pursue formal analysis of the OAuth 2.0 standard and proofs of security properties for all OAuth 2.0 flow types. Chen et al.~\cite{chen2014oauth} evaluate the use of OAuth in mobile applications and found 89 of 149 applications incorrectly implemented OAuth, thus making them vulnerable to attacks. In a 2012 field study of popular SSO systems, Wang et al.~\cite{wang2012signing} analysed SSO web traffic through the browser and identified 8 serious flaws in popular RPs and IdPs, allowing attackers to impersonate the victim user.

Mainka et al.~\cite{mainka2017sok} analyse the OpenID Connect protocol and identify security flaws similar to vulnerabilities found in other SSO protocols. They implement a fully-automated evaluation tool to identify implementation flaws in OpenID Connect libraries. Bai et al.~\cite{bai2013authscan} provide a tool to automatically identify security vulnerabilities in implementations of web authentication protocols including OAuth-based SSO. Yang et al.~\cite{yang2016model} propose an OAuth 2.0 security testing framework and automatically evaluate four IdPs (Facebook, Sina, Renren and Tencent Weibo) and 500 top-ranked web apps in US and China. Their empirical study reveals web apps that lack TLS protection for OAuth sessions leading to novel exploits.

Addressing challenges related to user awareness, AppCensus~\cite{appcensus} (cf.~\cite{reardon201950}) uses dynamic analysis to reveal privacy implications of granting data access to Android apps. More recently, Apple introduced \textit{privacy labels}~\cite{applePrivacyLabels} to highlight privacy practises to users of iOS apps. Narayanan et al.~\cite{narayanan2020dark} discuss \textit{dark pattern} designs in online services used to influence less-informed users into choices not in their best interest. Mathur et al.~\cite{mathur2019dark} investigate $\sim$11K shopping websites and find 1,818 instances of \textit{dark patterns} designed to increase user purchases. Felt et al.'s~\cite{felt2012android} user studies evaluate the effectiveness of Android permissions and find 20 of 24 participants were unaware or did not look at permission warnings. Unlike mobile apps where users are given only one set of permissions, SSO users often have the choice (although hidden) to login to a given RP with a less privacy-intrusive alternative so user awareness could significantly impact decisions.

Sun et al.~\cite{sun2011makes} empirically found users hesitant to adopt OpenID due to a lack of understanding and to concerns over releasing personal information. Many users held the misconception that their IdP credentials were shared with the RP. In 2012, Sun et al.~\cite{sun2012devil} also evaluated OAuth 2.0 implementations by three major IdPs (Facebook, Microsoft, Google) and explored 96 RP sites supporting the Facebook SSO. Results revealed several implementation decisions causing security concerns, including possible access token theft. Privacy implications discussed herein complement their work on security implications from identified vulnerabilities. Bonneau et al.~\cite{bonneau2012quest} surveyed 35 password-replacement schemes and found that compared to other schemes, federated SSO systems offer more benefits across various usability, deployability and security properties. Alaca et al.~\cite{alaca2020comparative} propose a framework to evaluate 14 web SSO schemes, including OAuth 2.0 and OpenID, and compare various properties including privacy benefits. They identify defining characteristics for each scheme and highlight priorities for stakeholders.

\section{Concluding Remarks}
\label{sec.concluding.remarks}
OAuth-based systems provide many benefits such as flexibility and convenience to SSO users. Services using OAuth benefit from reduced development costs related to outsourcing identity management.
When an RP supports multiple SSO logins, users must commit to an SSO option (and in many cases, complete the authentication) before finding what user data will be requested by the RP. This design means that users never find out what data would be requested by other SSO options, and consequently, are not fully informed about available choices on the RP site. Our empirical results reveal privacy practises where popular RPs request vastly different amounts of user data from different IdPs, with at least one option unquestionably more privacy-intrusive than others, similar to dark patterns found in website designs~\cite{mathur2019dark}~\cite{narayanan2020dark}.
SSO users are likely to make privacy decisions not in their best interest, due to the lack of information on available choices.

When granting RPs access to user data, users are not given information on the duration of the access. This lack of information, combined with an RP's ability to extend previously granted access without additional user involvement (Section~\ref{sec.framework.refreshTokens}), poses ongoing danger to user privacy. Further research is needed to mitigate risks related to allowing such continued access by RPs.

To the best of our knowledge, we offer the first in-depth analysis of OAuth-based SSO with a primary focus on user privacy as opposed to security. Based on the empirical work facilitated by our novel OAuthScope tool, we identify 8 areas to improve the privacy of OAuth. These improvements to OAuth's architecture will require effort and cooperation between IdPs, RPs, and specification authors. We hope that greater awareness by technically-savvy users and privacy enthusiasts, of the privacy implications identified through our work (Section~\ref{sec.privacy.implications}), may result in further attention to privacy violations, further community-based monitoring, and a more privacy-friendly OAuth-based SSO ecosystem.

\section*{Acknowledgment}
This work was partially funded by a grant through the Queen Elizabeth II Graduate Scholarship in Science and Technology (QEII-GSST) program. The second and third authors acknowledge funding from the Natural Sciences and Engineering Research Council of Canada (NSERC) through the Canada Research Chairs and Discovery Grant programs.

%The second author acknowledges funding from the Natural Sciences and Engineering Research Council of Canada (NSERC) for both his Canada Research Chair in Authentication and Computer Security, and a Discovery Grant. The third author acknowledges funding from NSERC for her Canada Research Chair in Human Oriented Computer Security.

\balance
\bibliographystyle{abbrv}
\bibliography{references}

\appendix

\section{OAuth 2.0 Authorization Request}
\label{app.authorization}
We provide an example authorization request discussed in Section~\ref{sec.oauthscope}. RPs specify OAuth 2.0 parameters in authorization requests to the IdP. A brief description is included for each parameter in the request~\cite{oauth2rfc}.
\newline

\texttt{HTTP GET /authorizationEndpoint?}
\par \texttt{response\_type=code}
\par \texttt{\&scope=email\%20profile}
\par \texttt{\&redirect\_uri=https\%3A\%2F\%2Fclient\%2Ecom\%2Fcb}
\par \texttt{\&client\_id=lp4qazfnh1}
\par \texttt{\&state=hnz3krb2mn}
\newline

\noindent
authorizationEndpoint: endpoint URI used by the RP for sending authorization requests to the IdP. \newline

\noindent
\texttt{response\_type:} specifies the OAuth flow type the RP intends to use with IdP.
\newline

\noindent
\texttt{scope:} a list of resources requested for access by the RP.
\newline

\noindent
\texttt{redirect\_uri:} user is redirected to this endpoint after completing interactions with the IdP. For security reasons, this value must match the endpoint registered with the IdP during RP's app registration.
\newline

\noindent
\texttt{client\_id:} a unique string issued to RP during registration.
\newline

\noindent
\texttt{state:} a unique (non-guessable) string generated by RP and included in the authorization request. The IdP returns the value when redirecting the user back to RP. To mitigate cross-side request forgery attacks, it must be ensured that the returned value is equal to value included in the initial request.

\section{OAuth 2.0 Implicit Flow}
\label{app.implicit}
\begin{figure*}[b]
    \centering
    \includegraphics[width=.8\textwidth]{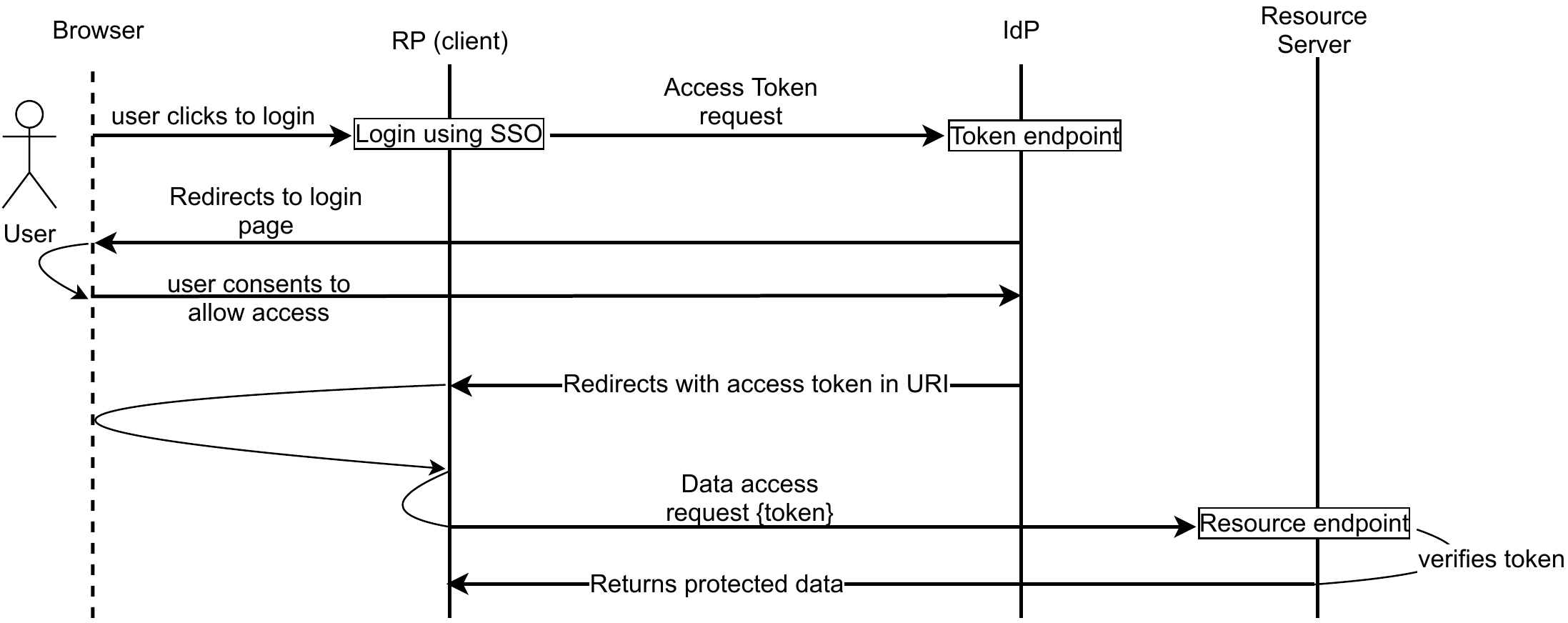}
    \caption{Procedure for the OAuth 2.0 Implicit flow (derived from ~\cite{oauth2rfc}).}
    \label{figImplicitFlow}
\end{figure*}

As an extension to background provided in Section~\ref{sec.framework}, Fig~\ref{figImplicitFlow} lists the process for the OAuth 2.0 implicit flow. Since the access token is returned to the RP in the redirection URI, it is vulnerable to token thefts from the user's browser.

\section{OAuthScope}
\label{app.screenshot}
\begin{figure*}[t]
    \centering
    \includegraphics[width=.8\textwidth]{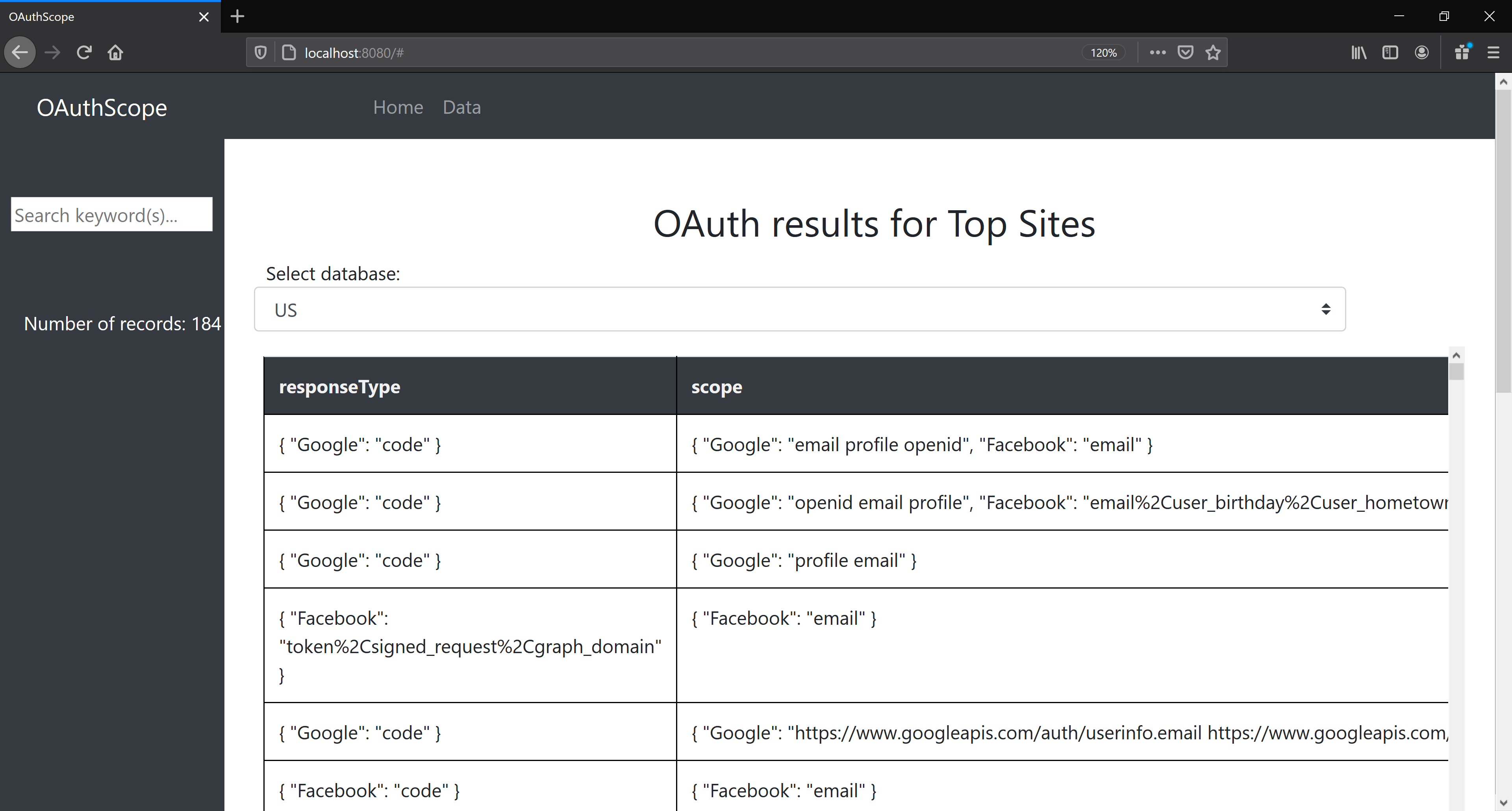}
    \caption{Screenshot of OAuthScope tool listing OAuth 2.0 parameters included in authorization requests from top US sites.}
    \label{figOauthscope}
\end{figure*}

Fig.~\ref{figOauthscope} is a screenshot of OAuthScope described in Section~\ref{sec.oauthscope} and lists identified OAuth 2.0 parameters for each RP. This UI is used for analysis of data collected by OAuthScope.

\end{document}